\documentclass[twocolumn,tighten]{aastex62}
\usepackage{amsmath,color}
\usepackage{array}
\usepackage{float}
\usepackage{graphicx}
\usepackage{subfigure}
\usepackage{natbib}

\shorttitle{Active Centaur photometry}
\shortauthors{Wong et~al.}

\begin{document}

\title{Photometry of active Centaurs: Colors of dormant active Centaur nuclei}

\author{Ian Wong}
\altaffiliation{\textit{51 Pegasi b} Fellow}
\affiliation{Department of Earth, Atmospheric, and Planetary Sciences, Massachusetts Institute of Technology,
Cambridge, MA 02139, USA}
\author{Aakash Mishra}
\affiliation{Davis Senior High School, Davis, CA 95616, USA}
\author{Michael E. Brown}
\affiliation{Division of Geological and Planetary Sciences, California Institute of Technology,
Pasadena, CA 91125, USA}
\correspondingauthor{Ian Wong}
\email{iwong@mit.edu}

\begin{abstract}
We present multiband photometric observations of nine Centaurs. Five of the targets are known active Centaurs (167P/CINEOS, 174P/Echeclus, P/2008 CL94, P/2011 S1, and C/2012 Q1), and the other four are inactive Centaurs belonging to the redder of the two known color subpopulations (83982 Crantor, 121725 Aphidas, 250112 2002 KY14, and 281371 2008 FC76). We measure the optical colors of eight targets and carry out a search for cometary activity. In addition to the four inactive Centaurs, three of the five active Centaurs showed no signs of activity at the time of observation, yielding the first published color measurements of the bare nuclei of 167P and P/2008 CL94 without possible coma contamination. Activity was detected on P/2011 S1 and C/2012 Q1, yielding relatively high estimated mass loss rates of $140\pm20$ and $250\pm40$~kg/s, respectively. The colors of the dormant nuclei are consistent with the previously-published colors, indicating that any effect of non-geometric scattering from Centaur dust or blanketing debris on the measured colors is minimal.  The results of our observations are discussed in the context of the cause of Centaur activity and the color distributions of active and inactive Centaurs. We suggest that the relative paucity of red Centaurs with low-perihelion orbits may not be directly due to the blanketing of the surface by unweathered particulates, but could instead be a result of the higher levels of thermal processing on low-perihelion Centaurs in general.

\end{abstract}

\section{Introduction}

Centaurs are a population of minor bodies located in the middle-outer Solar System and are conventionally defined to be objects with orbital semimajor axes and perihelion distances between those of Jupiter and Neptune. These objects originated farther out in the Kuiper Belt and were scattered inward onto unstable orbits with dynamical lifetimes around 10~Myr \citep{tiscareno}. Their smaller heliocentric distances have made Centaurs ideal proxies for studying the more distant Kuiper belt object (KBO) population from which they are sourced. More broadly, understanding the composition of Centaurs and KBOs may provide crucial constraints on the properties of the outer regions of the primordial protoplanetary disk, with major implications for formation and evolution models of the Solar System.

Spectroscopic and photometric observations have revealed that Centaurs are generally characterized by redder-than-solar colors and very low albedos. Most notably, the optical color distribution of Centaurs is strongly bimodal, dividing the population into two groups: the gray and red Centaurs \citep[e.g.,][]{peixinho2003,romanishin}. Until recently, the origin of this bifurcation in color remained an open question. By analyzing the colors of KBOs, several authors suggested that they are likewise bimodal in color \citep{fraserbrown,peixinho2012}. However, at the time, well-measured KBO colors were relatively few and largely obtained through targeted observations. As such, the observational biases were difficult to quantify, and our understanding of the underlying color distribution of KBOs was uncertain. 

Recent surveys of small KBOs in the same size range as Centaurs ($d<300$~km, assuming a visible albedo of $\sim$0.1) have definitively resolved the question of Centaur color bimodality \citep[e.g.,][]{wong2017,ossos}. These systematic photometric observations have revealed that all dynamical KBO classes except the cold classicals also display a robust color bimodality. Moreover, the measured color centers of the two color subpopulations are identical to the ones observed in the gray and red Centaurs. This indicates that the observed optical color bimodality among Centaurs is an inherited feature from their source population --- the KBOs.

A subset of Centaurs of particular interest is the active Centaurs. These unusual objects have been observed to display comet-like behavior at times, expelling gas and dust to form a dispersed coma of material around a central nucleus. The activity of these Centaurs provides a unique window into their chemical and physical characteristics and may offer crucial insight into the composition and evolutionary history of outer solar system minor bodies in general. A notable property of active Centaurs is that their measured color distribution has been shown to differ from that of the overall Centaur population. Specifically, active Centaurs appear to generally lack the very red surface colors found among the red Centaur subpopulation \citep[e.g.,][]{jewitt,jewitt2015}.

There are three salient open questions concerning active Centaurs. First, there is the question of whether the activity itself is systematically affecting the measured colors of active Centaurs. The vast majority of the color measurements of active Centaurs in the published literature have been obtained when the bodies were active. Non-geometric scattering off optically small coma particles ejected from an active Centaur could introduce a relatively blue component to the photometry, as has been observed, for example, on some short-period comets \citep[e.g.,][]{ahearn}. The second question is whether both primordially gray and red KBOs can become active when scattered inward onto Centaur orbits. It is possible that the cometary activity of these objects could be blanketing the surface with pristine material from below the highly-irradiated outer layers \citep[e.g.,][]{delsanti,jewitt,jewitt2015}, which may have a distinct color and thus obscure the original color of the body prior to activity.  Alternatively, the apparent color-activity dependence could be reflective of different physical properties, which would suggest that the gray and red Centaurs, and by extension gray and red KBOs, may have formed in separate regions of the protoplanetary disk. This second question has been nominally settled by the recent discovery of the first active red Centaur \citep{mazzotta}, though the aforementioned discrepancy between the overall color distributions of active and inactive Centaurs remains unaddressed.

\begin{deluxetable*}{ccccccccc}[t!]
\tabletypesize{\scriptsize}
\tablecaption{
    Observation Details and Photometry\label{tab:obs}
}
\tablecolumns{9}
\tablehead{
    \colhead{Object} &
    \colhead{UT Date} &
    \colhead{B,V,R $n_{\mathrm{exp}}$\tablenotemark{a}} &
    \colhead{B,V,R $t_{\mathrm{exp}}$ (s)\tablenotemark{a}} &
    \colhead{$r$ (px)\tablenotemark{b}} &
    \colhead{$R_h$ (AU)\tablenotemark{c}} &
    \colhead{$R$}  &
    \colhead{$B-V$}  &
    \colhead{$V-R$} 
}
\startdata
83982 Crantor & 2017 Aug 21 & 15,4,4 & 1500,480,400 & 4.5 & 18.46 & $21.20\pm0.03$  & $1.02\pm0.08$ & $0.78\pm0.06$ \\
121725 Aphidas & 2017 Feb 15 & 15,7,7 & 2250,1140,720 & 3.0 & 25.87 & $22.59\pm0.06$  & $1.02\pm0.15$ & $0.63\pm0.10$ \\
250112 2002 KY14 & 2017 Feb 16 & 0,1,3 & 0,120,600 & 3.5 & 11.91 & $20.39\pm0.01$  & --- & $0.67\pm0.03$\\
281371 2008 FC76 & 2017 Feb 15 & 5,4,2 & 180,105,70 & 3.0 & 10.65 & $19.77\pm0.02$  & $0.93\pm0.04$ & $0.65\pm0.03$\\
'' & 2017 Feb 16 & 0,1,3 & 0,120,420 & 3.0 & 10.65 & $19.72\pm0.01$  & --- & $0.69\pm0.03$\\
167P/CINEOS   & 2017 Feb 15 & 9,5,15 & 880,150,1200 & 3.0 & 17.15 &  $22.13\pm0.03$  & $0.92\pm0.15$ &   $0.43\pm0.12$\\
'' & 2017 Aug 21 & 20,8,6 & 3000,960,720 & 3.5 & 17.35&  $22.17\pm0.03$  & $0.80\pm0.06$ & $0.51\pm0.06$\\
174P/Echeclus & 2017 Aug 21 & 2,1,4 & 280,120,240 & 4.5 & 7.03 & $18.31\pm0.01$  & $0.93\pm0.01$ & $0.48\pm0.01$ \\
P/2008 CL94 & 2017 Aug 21 & 2,2,2 & 200,120,120 & 4.5 & 6.33 &  $20.27\pm0.02$  & $0.89\pm0.04$ & $0.47\pm0.03$ \\
P/2011 S1 & 2017 Feb 16 & 0,0,4 & 0,0,780 &  5.5 & 7.36 & $21.88\pm0.05$ & --- & --- \\
C/2012 Q1 & 2017 Aug 21 & 8,4,4 & 960,320,480 & 4.5 & 12.79 & $22.21\pm0.04$  & $1.04\pm0.14$ & $0.58\pm0.10$\\
\enddata
\tablenotetext{a}{Number of individual exposures and total exposure time in the B, V, and R filters, respectively.}
\tablenotetext{b}{Radius of aperture used for photometric extraction. The pixel scale is $0.363''$.}
\tablenotetext{c}{Heliocentric distance at the time of observation.}
\end{deluxetable*}

Lastly, the causes of Centaur activity and its effects on the surface properties of these objects are not fully understood. The orbital perihelia of active Centaurs are significantly smaller than those of inactive Centaurs \citep{jewitt}, suggesting that the activity is primarily dependent on surface temperature. The heliocentric distances at which active Centaurs have been observed to display coma are consistent with the activity being triggered by the crystallization of amorphous ice and the concomitant release of fine-particle dust and trapped volatile gases, such as carbon monoxide \citep[e.g.,][]{capria,jewitt,wierzchos}. However, the importance of secondary processes affecting the surface chemistry remains unknown. 

As part of the ongoing effort to better understand the nature of active Centaurs, we carried out a photometric study of nine Centaurs. These observations targeted in particular active Centaurs that have passed perihelion and are at larger heliocentric distances than when they were previously observed to be active. The main objective was to look for cessations in activity and obtain color measurements of dormant active Centaur nuclei without the possible contamination from coma material. A complementary goal was to search for the onset of activity on known red Centaurs. Monitoring the activity of Centaurs provides strong constraints on the causes of activity, with major implications for our understanding of the physical properties and chemical composition of these objects.

\section{Observations and Data Analysis}\label{sec:obs}

We carried out photometric observations of Centaurs on three nights in 2017 --- February 15--16 and August 21. The targets were imaged in three bands --- B (390--490~nm), V (505--595~nm), and R (570--690~nm) --- using the Large Format Camera (LFC) instrument on the 200-inch Hale Telescope at Palomar Observatory. The science detector in LFC consists of an array of six 2048$\times$4096 CCDs with a pixel scale of 0.18$''$. We applied 2$\times$2 binning to reduce readout time, resulting in an effective pixel scale of 0.363$''$. Bias frames and dome flats were acquired at the beginning of each night prior to science observations. Exposure times were calculated so as to ensure a per-filter signal-to-noise of at least 50, assuming typical observing conditions at Palomar; imaging in R-band, where Centaurs are intrinsically brighter, was extended to increase sensitivity to activity, yielding signal-to-noise values exceeding 100 in many cases.

For all but the brightest targets, observations in each filter were split across multiple exposures. Filters were cycled sequentially to minimize the effect of possible rotational brightness modulation on the color measurements. Several of our targets have published rotational lightcurve measurements: all of them have periods longer than 7~hours and peak-to-peak photometric amplitudes of less than 0.25~mag, while the overall mean Centaur rotational period from lightcurve surveys is around 9~hours \citep{duffard}. For our observations, the maximum time baseline for consecutive B-V-R exposures is under 20~minutes. It follows that the expected contribution of rotational lightcurve modulation to individual color measurements is small. This methodology is empirically validated when comparing independent $B-V$ and $V-R$ color measurements we obtained for the same object --- in all cases, the measurements are self-consistent at better than the 1.5$\sigma$ level. Therefore, there is no statistically significant effect of the objects’ rotation on our color measurements.

Observations during the two consecutive February nights were plagued by intermittent cloud cover, high sky background, and non-photometric conditions. Only several hours worth of observation on 2017 Feb 15 produced data of sufficient quality to yield reliable colors. Conditions on 2017 Feb 16 were significantly better, and we dedicated that evening's observations to long exposures, primarily in R-band, in search of activity. As a result, several of the Centaurs imaged on the 2017 February nights do not have $B-V$ and $V-R$ color measurements. For the exposures from the February nights that we used in our analysis, the seeing ranged from 1.0$''$ to 2.0$''$. Conditions on 2017 Aug 21 were good, with clear skies throughout the night. Seeing during the final night mostly varied between 1.2$''$ and 1.8$''$.

The images were run through a pipeline that handles data reduction, image processing, and photometric calibration using standard techniques for moving object photometric observations. Following bias-subtraction and flat-fielding, we detected bright sources in each image and automatically matched them with stars in the Pan-STARRS DR1 catalog \citep{flewelling} to produce an astrometric solution. Next, the Pan-STARRS catalog magnitudes were converted to B,V,R magnitudes using the empirical conversions calculated by \citet{tonry}, and the photometric zeropoint was calculated for each image in the respective band. The instantaneous position of each target at the time of exposure was determined by querying the JPL Horizons database, from which the corresponding source on the image was identified and its apparent magnitude measured. We utilized fixed circular apertures for photometric extraction ranging in size from 3 to 8 pixels in radius, choosing the optimal aperture radius for each object that minimized the resultant uncertainty in the color measurements. Exposures in which the target was situated in close proximity to other sources or overlapped with hot pixels and/or cosmic rays were removed from our analysis. The details of our observations are listed in Table~\ref{tab:obs}.

To search for Centaur activity, we constructed surface brightness profiles (SBPs) using multi-aperture R-band photometry for the target and 3--5 non-saturated bright stars in each image. The flux density within concentric anulli of one-pixel width was computed and plotted as a function of distance from the centroid. The SBPs of stellar sources were averaged together and renormalized to line up with the SBP of the target at low separation distances. This procedure was carried out for each target and the resulting plots inspected by eye to look for significant excess flux in the target's SBP.

\section{Results}\label{sec:results}

Across the three nights of observation, we obtained high-quality photometry of 9 Centaurs. Five of the targets are known active Centaurs --- 167P/CINEOS, 174P/Echeclus, P/2008 CL94, P/2011 S1, and C/2012 Q1; the remaining four are VR Centaurs --- 83982 Crantor, 121725 Aphidas, 250112 2002 KY14, and 281371 2008 FC76. The calculated apparent $R$ magnitudes and $B-V$ and $V-R$ colors are listed in Table~\ref{tab:obs} along with the heliocentric distances of each object at the time of observation. The objects 167P and 281371 were observed on two nights, and the individual color measurements are statistically consistent to within 1$\sigma$.

\begin{figure*}[t!]
\centering
\includegraphics[width=0.8\linewidth]{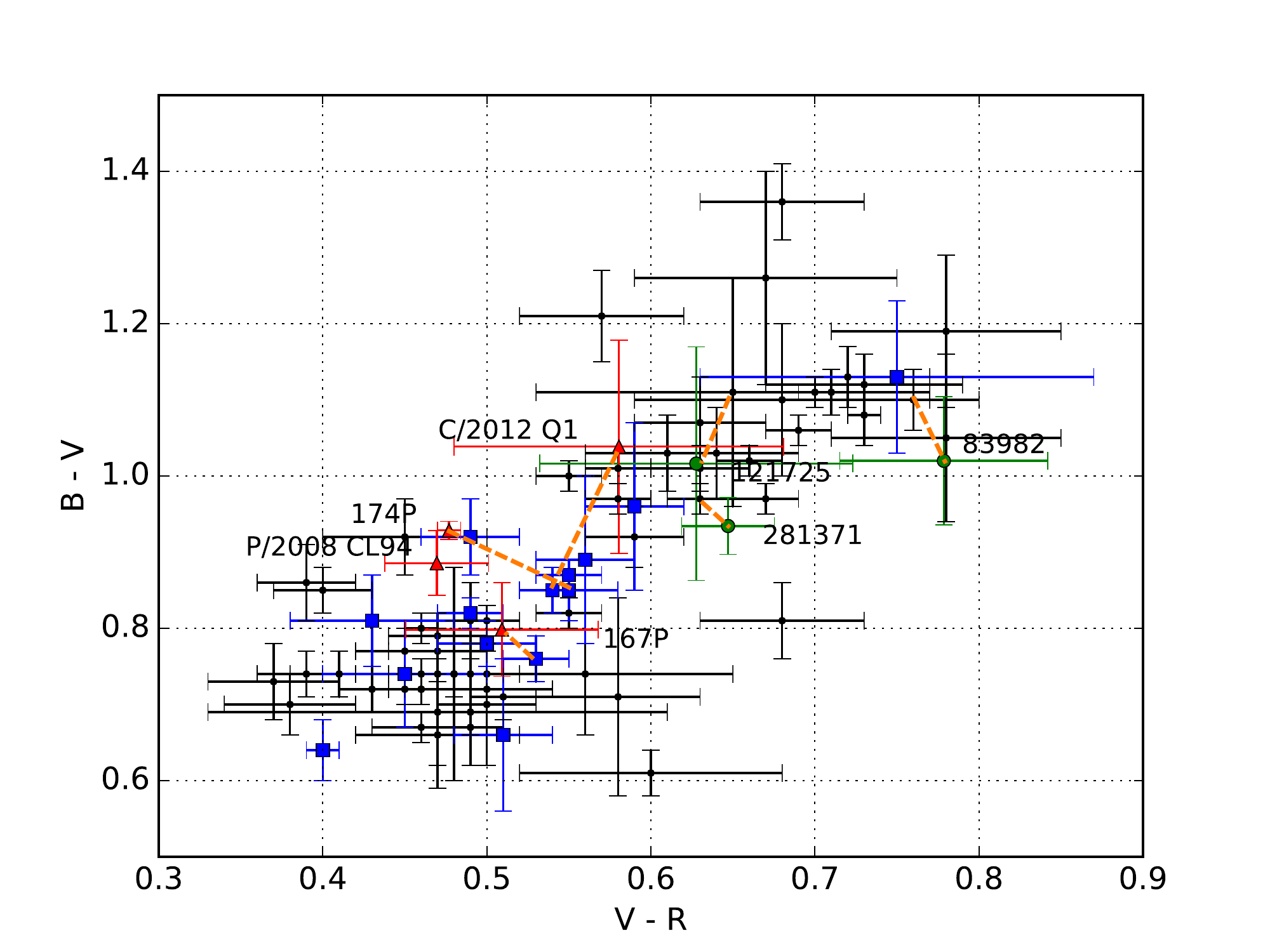}
\caption{Color-color plot of Centaurs. Colors from our photometric observations of active and inactive Centaurs are indicated by red triangles and green circles, respectively, and are labeled by their designations. The colors of inactive Centaurs from the literature \citep{jewitt2015,peixinho2015,tegler2016} are indicated by black dots, while previously-published colors of active Centaurs \citep{jewitt,mazzotta2014,jewitt2015,peixinho2015,tegler2016,mazzotta} are indicated by blue squares. For 167P, only the color measured from our 2017 Aug 21 observations is shown in red. Orange dashed lines link our color values with the most precise published values in the literature for the same objects, when available. The bimodality of the inactive Centaur color distribution is discernible. Active Centaurs with published colors in the literature are largely consistent with the less-red of the two subpopulations, i.e., gray Centaurs, with the marked exception of the recently characterized red active Centaur 523676. The known active Centaurs in our sample have $B-V$ and $V-R$ colors that are consistent with the gray Centaurs. The inactive Centaurs in our sample belong to the red Centaur subpopulation.}
\label{fig:colors}
\end{figure*}

We calculated both $B-V$ and $V-R$ colors for seven objects: 83982, 121725, 281371, 167P, 174P, P/2008 CL94, and C/2012 Q1. Their colors are plotted in Figure~\ref{fig:colors}. We also include all other Centaur colors compiled from the published literature \citep{jewitt,mazzotta2014,jewitt2015,peixinho2015,tegler2016,mazzotta}. For an object with multiple observations, we take the most precise published color measurement. Some active Centaurs have photometry from multiple apertures listed; in those cases, the colors from the smallest aperture are shown, as they are most reflective of the nucleus color. Black dots indicate inactive Centaurs, while blue squares denote known active Centaurs not included in our observational sample. Our color measurements of active and inactive Centaurs are plotted with red triangles and green circles, respectively. Orange dashed lines link our color measurements with the literature values --- all objects have color values that are consistent with previously-published values to within $2\sigma$, with most lying well within $1\sigma$ of the literature values.

From the plot, it is evident that the color distribution of inactive Centaurs is strongly bimodal, as has been pointed out by many authors \citep[e.g.,][]{peixinho2003,peixinho2015,jewitt2015,tegler2016,romanishin}, dividing the population into the gray and red Centaur subpopulations. The inactive red Centaurs in our target list --- 83982, 121725, 281371 --- were measured to have $B-V$ and $V-R$ colors characteristic of the red subpopulation. For 250112, only the $V-R$ color was measured; it is consistent with the $V-R$ colors of the other red Centaurs as well as literature values.

Turning to the active Centaurs, no bimodality is evident. Most of these objects, including the four dormant active Centaurs in our observational sample, have colors consistent with the gray Centaur subpopulation, while a few have colors that lie intermediate between the two color modes in the inactive Centaur color distribution. A notable outlier is 523676 2013 UL10, which has measured colors that lie solidly within the red Centaur subpopulation \citep{mazzotta}. We note that our measurement of C/2012 Q1 is highly uncertain, but statistically consistent with the previous measurement by \citet{jewitt2015}, which makes the object consistent with the gray Centaur subpopulation.

\begin{figure*}[t!]
\includegraphics[width=0.5\linewidth]{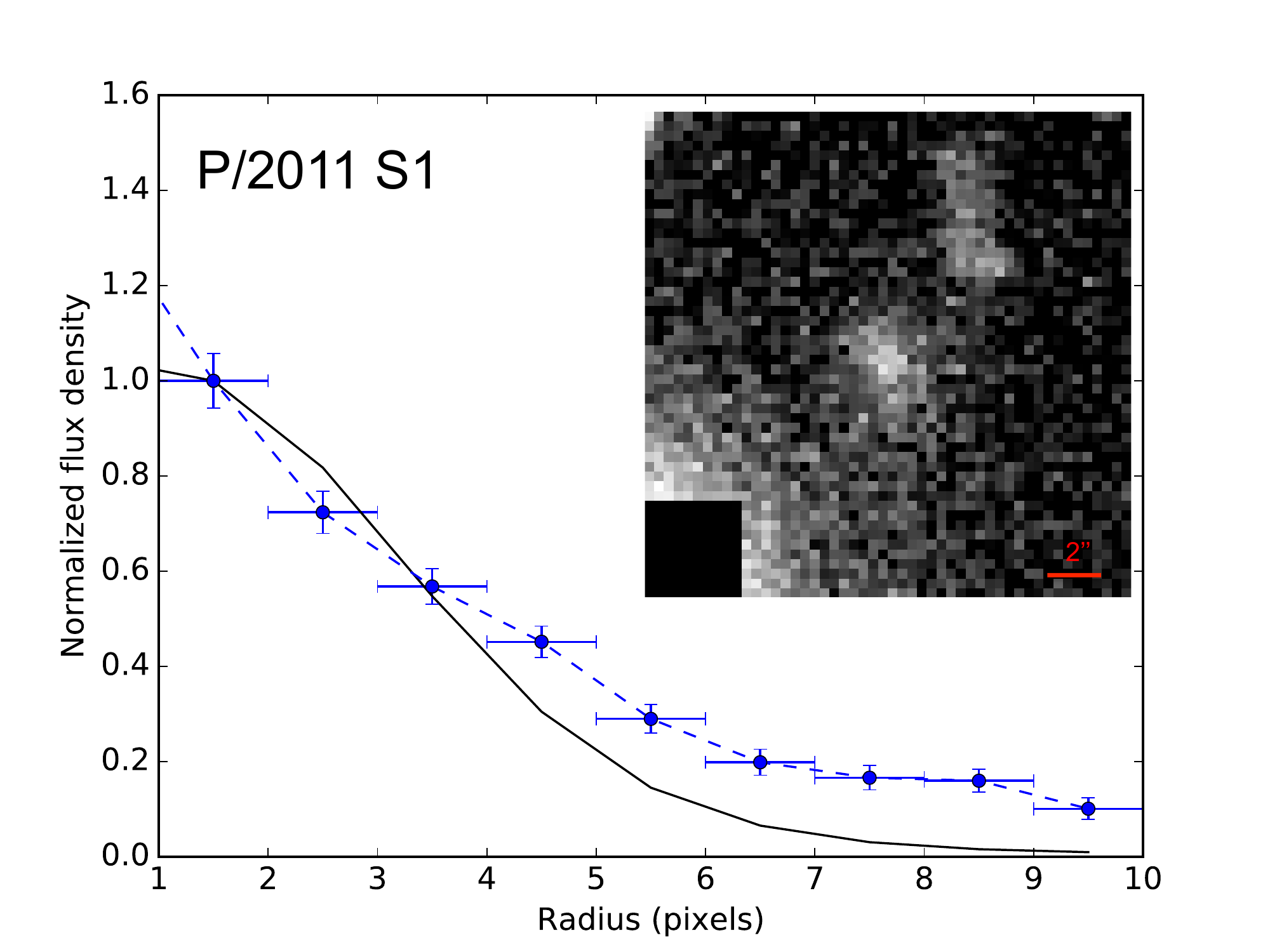}
\includegraphics[width=0.5\linewidth]{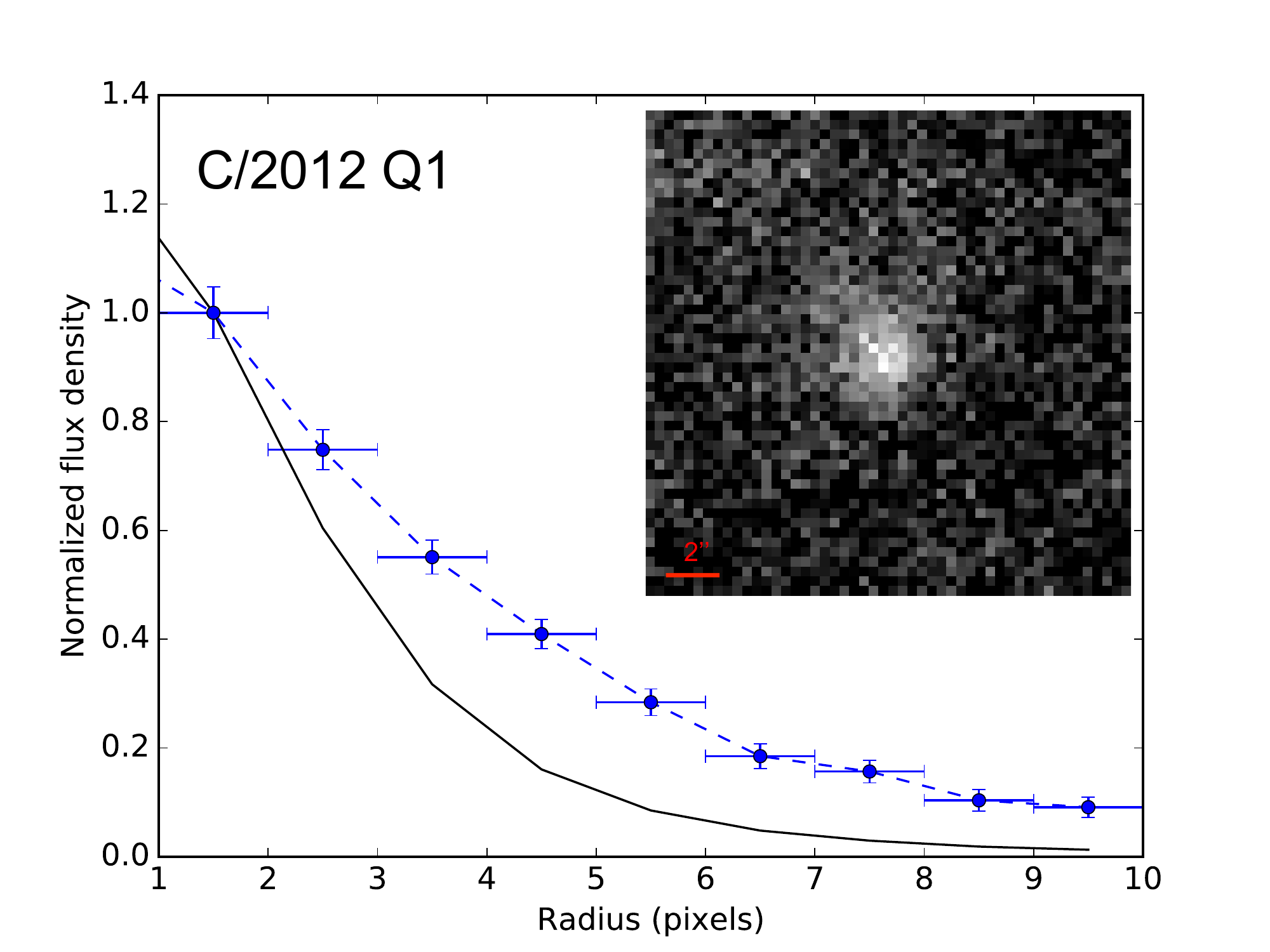}
\caption{Plot displaying the surface brightness profiles (SBPs) of P/2011 S1 (left) and C/2012 Q1 (right) in blue, alongside the averaged SBPs of bright field stars in black. Stacked 50$\times$50~px R-band images centered on the targets are also provided. The SBPs have been arbitrarily normalized such that the flux density between 1 and 2 pixels from the centroid is unity. The excess flux density in the wings of each Centaur's point-spread function is evident, indicating scattering from coma particulates.}
\label{fig:SPB}
\end{figure*}

From careful inspection of both stacked images and the derived SBPs for each target, we detected activity around P/2011 S1 and C/2012 Q1 at the time of our observations. The averaged SBPs of these two targets are plotted alongside the averaged SBPs of field stars in Figure~\ref{fig:SPB}. Coadded R-band images, centered on the targets, are also provided. We have arbitrarily renormalized all SBPs to unity between 1 and 2 pixels from the centroid. From the figure, it is evident that the SBPs of the two active Centaurs show significant deviation from the point source SBPs, indicating reflection off coma material. 

The other active Centaurs in our sample --- 167P, 174P, and P/2008 CL94 --- have SBPs that are consistent with a point source and thus were dormant at the time of our observations. It follows that their measured colors in Table~\ref{tab:obs} can be interpreted as the uncontaminated colors of their nuclei. For 167P and P/2008 CL94, our observations provided the first measured colors during a period of inactivity. None of the inactive Centaurs in our sample showed evidence for coma.

In order to study the activity level of our Centaur targets in more detail and place constraints on the coma brightnesses and dust production rates, we carried out an analogous analysis to the ones presented in several previous active Centaur publications \citep[e.g.,][]{jewitt,mazzotta2011,mazzotta2014}. The magnitude of the coma in R-band is calculated using the following expression:
\begin{equation}\label{coma}m_{c} = -2.5\log_{10}\left\lbrace \eta \times [f(r_{2})-f(r_{1})]\right\rbrace+m_{0},\end{equation}
where $m_{0}$ is the photometric zeropoint computed from photometric calibration of each R-band image, and $f(r)$ is the flux contained within an aperture $r$. For the coma measurements presented in this paper, we chose $r_{1}=6$~px and $r_{2}=10$~px, corresponding to radii of 2.2 and 3.6$''$. The factor $\eta$ is a correction factor to remove the flux contribution from the wings of the nucleus' point-spread function (PSF) lying within the annulus. The value of $\eta$ was estimated from computing the fraction of each field star's PSF that is contained within the annulus and averaging. In our images, $\eta$ varied from $\sim$0.80 to 0.98 due to the relatively large changes in seeing throughout the nights of observation (ranging from $0.9''$ to over $2.0''$).

From here, the coma cross section is given by \citep{russel1916}
\begin{equation}\label{eq:cross}C_{c} = 2.25\times10^{22} \frac{\pi R^2\Delta^2 10^{-0.4(m_{c}-m_{\Sun})}}{p_{R}\Phi(\alpha)},\end{equation}
where $R$ and $\Delta$ are the heliocentric and geocentric distances (in AU), and $m_{\Sun}=-27.15$ is the R-band apparent magnitude of the Sun. For consistency with previous active Centaur studies \citep[e.g.,][]{romanishin} and ease of comparison with other activity estimates in the literature \citep[e.g.,][]{jewitt,mazzotta2011,mazzotta2014,mazzotta}, we set the geometric albedo $p_{R}$ to 0.1. Similarly for consistency with previous studies, we use a generic form for the phase function $\Phi(\alpha)=10^{-0.4\beta\alpha}$, where $\alpha$ is the phase angle in degrees, and we assume $\beta=0.1$~mag/degree.

The total coma mass is estimated in the same way as in the \citet{jewitt} and \citet{mazzotta} analyses: $M_{c}=\frac{4}{3}\rho \bar{a} C_{c}$, where we set the average size of the coma dust to $\bar{a}=30$~$\mu$m and assume $\rho=1000$~kg/m$^3$. It follows that the mass loss rate is simply the total coma mass $M_{c}$ divided by the average residence time of coma particles within the sky-projected annular region between $r_{1}$ and  $r_{2}$. We assume that the dust particles follow the heliocentric distance dependent trend in outflow velocity that was derived from observations of C/Hale-Bopp \citep{biver2002}: $v(R) = 550~\mathrm{m/s}\times(5~\mathrm{AU}/R_h)^{1/4}$. The residence time $\tau$ is thus
\begin{equation}\tau(R) = \frac{\Delta\times(r_{2}-r_{1})}{550~\mathrm{m/s}}\left(\frac{R_h}{5~\mathrm{AU}}\right)^{1/4},\end{equation}
where $r$ is in arcseconds, and $\Delta$ is in meters.

The results of our calculations are listed in Table~\ref{tab:act}. For objects with no discernible activity, we derive 3$\sigma$ upper limits based on the standard deviation of background scatter. The relative background level in our observations of 121725 was too high and the object too faint to allow for even modest constraints on activity level, and we do not report the resultant upper limits.

\begin{deluxetable}{cccc}
\tablewidth{0pc}
\tabletypesize{\scriptsize}
\tablecaption{
    Centaur Activity
    \label{tab:act}
}
\tablehead{
    \colhead{Object} &
    \colhead{$m_c$\textsuperscript{a}}                     &
    \colhead{$M_c$ ($10^6$ kg)\textsuperscript{b}}  &
    \colhead{$\dot{M_c}$ (kg/s)\textsuperscript{c}} 
}
\startdata
83982 Crantor & $>26.1$ & $<1.6$ & $<30$\\
250112 2002 KY14 & $>26.4$ & $<0.3$ & $<10$ \\
281371 2008 FC76 & $>25.9$ & $<0.3$ & $<10$ \\
167P/CINEOS\textsuperscript{d} &  $>26.2$ & $<1.5$ & $<30$\\
''\textsuperscript{d} & $>26.0$ & $<1.8$ & $<40$\\
174P/Echeclus & $>25.7$ & $<0.09$ & $<7$ \\
P/2008 CL94 & $>26.1$ & $<0.03$ & $<3$  \\
P/2011 S1 & $22.2\pm0.1$ & $2.0\pm0.2$ & $140\pm20$ \\
C/2012 Q1  & $22.7\pm0.2$ & $7.5\pm1.2$ & $250\pm40$ 
\enddata
\tablenotetext{a}{Coma magnitude between 6 and 10 pixels (2.2--3.6$''$) from the center of the object. For Centaurs with no discernible activity, $3\sigma$ upper limits are given.}
\tablenotetext{b}{Coma mass.}
\tablenotetext{c}{Mass loss rate.}
\tablenotetext{d}{These two measurements are derived from our observations on 2017 Feb 15 and 2017 Aug 21, respectively.}
\end{deluxetable}

In the following, we briefly summarize the previous observational studies of the active Centaurs in our sample. A diagram illustrating the activity histories of these objects is shown in Figure~\ref{fig:activity}.

\subsection{167P/CINEOS}\label{subsec:167P}
167P is a distant active Centaur, coming to perihelion beyond the orbit of Saturn ($q=11.8$~AU). This object was observed to display a faint coma on 2004 Oct 10, when the object was at a heliocentric distance of $R_h=12.23$~AU \citep{jewitt}. During that epoch, 167P had an R-band magnitude of $20.69\pm0.02$, with measured colors of $B-V=0.80\pm0.03$ and $V-R=0.49\pm0.10$. The mass loss rate was estimated to be 24~kg/s. Since then, the object has continued on the outward branch of its orbit. Activity was detected on 2005 Jun 7--8 \citep[$R_h=12.41$~AU;][]{romanishin2005}, 2010 Sep 10 ($R_h=14.42$~AU), and 2012 Oct 13 \citep[$R_h=15.33$~AU;][]{jewitt2015}. The measured colors within an $4.1''$-radius aperture during 2010 Sep 10 were $B-V=0.80\pm0.04$ and $V-R=0.57\pm0.03$, while the colors within a larger $6.8''$ aperture on 2012 Oct 13 were notably bluer -- $B-V=0.68\pm0.05$ and $V-R=0.34\pm0.05$ -- indicating that the coma material appears significantly bluer than the nucleus. 

Our observations of 167P from 2017 Feb 15 ($R_h=17.15$~AU) and 2017 Aug 21 ($R_h=17.35$~AU) revealed no discernible activity. We note, however, that even though inspection of the images and surface brightness profiles did not find visible signs of activity, given the quality of our observations and the much larger heliocentric distances of the target at the time, the $3\sigma$ upper limits on mass loss rate (30--40~kg/s) are still nominally consistent with the estimated value in \citet{jewitt}. Our measured colors within apertures of $1.1''$ and $1.3''$ are consistent with each other and statistically consistent with those presented by \citet{jewitt} and \citet{jewitt2015} (for the smaller aperture) at better than the 1$\sigma$ level.

\subsection{174P/Echeclus}\label{subsec:174P}
174P is one of the most well-studied Centaurs and has an activity history that is characterized by several distinct outbursts. Pre-discovery Spacewatch images of 174P showed that a discernible coma was present in 2000 January ($R=15.4$~AU) when the object was near aphelion \citep{wierzchos}. Analysis of subsequent observations in 2001--2003 ($R_h$ = 13.6--14.5~AU) yielded a rotation period of $26.802\pm0.042$~hr, and the first color measurements ($B-V = 0.76\pm0.15$ and $V-R = 0.51\pm0.09$), but did not detect any coma \citep{rousselot2005}. These color measurements obtained during a period of inactivity are consistent with the values we obtained from our observations. The first detected outburst occurred in late 2005 ($R_h=13.1$~AU), during which the 174P's brightness increased by 7~mag. \citep[e.g.,][]{choi2006}. Subsequent observations in 2006 February--April ($R_h$ = 12.9--13.0~AU) revealed that the coma had a complicated and highly-extended structure, discernible out to $2'$ from the object center \citep{tegler2006,bauer2008,rousselot2008,jewitt}. Peculiarly, the nexus of the coma was a secondary component separated from and in motion around the primary.  Explanations for this unusual coma include fragmentation of the primary due to either explosive outgassing or an impact \citep[e.g.,][]{rousselot2008,fernandez2009}. Multiband photometry by \citet{bauer2008} produced color measurements of $B-V=0.77  \pm0.06$; $V-R=0.51\pm0.04$ and no significant variation in color with aperture size; in particular, comparison between colors computed for a circular aperture centered on the nucleus and for an extended wide-separation annulus showed no significant difference. \citet{rousselot2008} presented similar color measurements. Mass loss rate estimates during this first observed outburst ranged from 100 to 400~kg/s.

Subsequent documented outburst events were observed in 2011 May--July \citep[$R_h$ = 8.4--8.5~AU;][]{jaeger2011,rousselot2015} and  2016 August--September \citep[$R_h$ = 6.3~AU;][]{miles2016}, with the latter outburst ending in early 2016 October. Colors obtained by \citet{rousselot2015} during the second outburst are consistent with measurements obtained during the first outburst. During the second and third outbursts, no secondary components were detected. Mass loss rate estimates were consistently lower than values from the first outburst, lying in the tens of kg/s. The past episodes of activity were separated by null detections of discernible coma material, when the object was at heliocentric distances of 13.6--14.5, 11.3--12.2, 5.9--6.6, and 7.0~AU \citep[][as well as our measurements]{bauer2003,rousselot2005,lorin2007,rousselot2008,choi2008,rousselot2015,tegler2016}. 

\begin{figure}[t!]
\includegraphics[width=\linewidth]{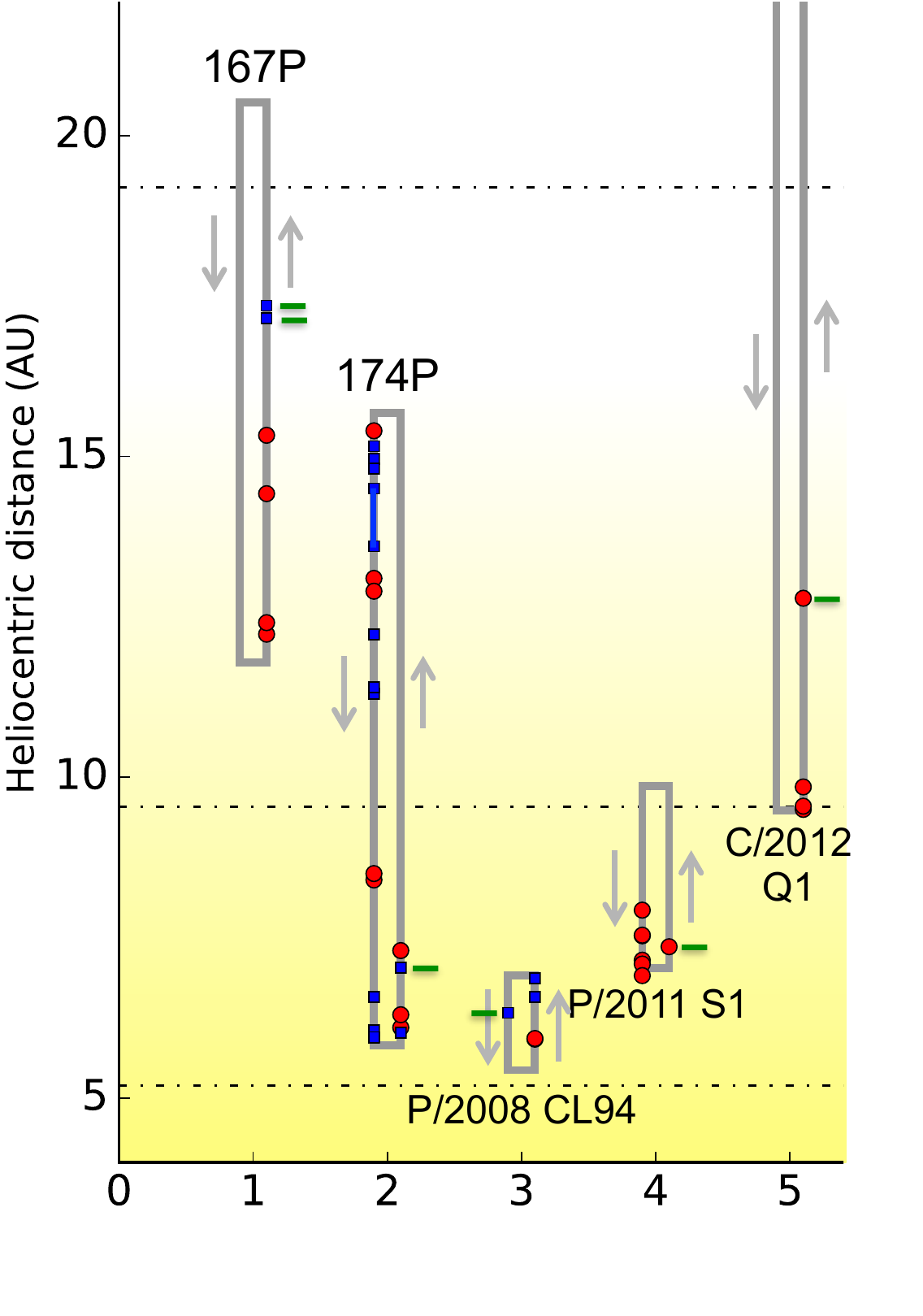}
\caption{A summary of the published activity history for the five active Centaurs in our sample (see Sections~\ref{subsec:167P}--\ref{subsec:C2012}). The gray circuits represent the objects' orbits in the counterclockwise sense. The orbital semi-major axes of Jupiter, Saturn, and Uranus are represented by horizontal lines. Red points indicate epochs when an object was observed to be active, while blue squares indicate periods of observed inactivity. Green dashes mark the datapoints corresponding to our observations. The yellow shaded region represents the range of heliocentric distances at which amorphous-to-crystalline water ice conversion is predicted to occur; at the largest heliocentric distances within this range, high obliquity is needed in order for this process to be an effective driver of cometary activity.}
\label{fig:activity}
\end{figure}

Spectroscopic observations of 174P near its perihelion passage in 2016 May--June ($R_h=6.1$~AU), just prior to the third documented outburst, yielded a $3.6\sigma$ detection of CO emission, indicating an active gas production rate of $(7.7\pm3.3)\times´10^{26}~\mathrm{mol/s}$ \citep{wierzchos}. This makes 174P only the third Centaur with detected CO, after the extremely active 29P/Schwassmann–Wachmann 1 and 2060 Chiron \citep[see][and references therein]{womack2017}. An independent set of spectroscopic observations before (2014 Aug 3; $R_h=5.95$~AU) and after  (2016 Oct 7--8; $R_h=6.34$~AU) this third outburst found no significant changes to the spectral gradient from pre- to post-outburst \citep{seccull}. Notably, analysis of the coma spectra yielded a blue spectral gradient of $−7.7\pm0.6$~\%/1000~\AA, consistent with co-eval annular aperture photometry that indicated coma colors that were significantly bluer than the near-nucleus measurements.

Most recently, a fourth large outburst was detected by amateur astronomers in 2017 December\footnote{http://britastro.org/node/11988}, when the object was 7.3~AU from the Sun. No further   observations have been documented as of the writing of this paper.

\subsection{P/2008 CL94}\label{subsec:P2008}
The orbit of P/2008 CL94 has perihelion and aphelion distances of $q=5.43$~AU and $Q = 6.91$~AU. This object has been alternatively classified as a Jupiter-family comet (JFC) or a Centaur-JFC transition object \citep[e.g.,][]{horner2003,kulyk}. Activity was observed on 2009 Mar 17--18 \citep[$R_h=5.92$~AU;][]{scotti2009}. Photometric and spectroscopic observations on 2009 Apr 1 ($R_h=5.93$~AU) revealed cometary activity, an R-band magnitude of $19.64\pm0.01$, and a featureless visible spectrum with a red continuum slope of $S = 2.0\pm0.4$~\%/1000~\AA~\citep{kulyk}. Later observations on 2011 Jul 7 ($R_h=6.58$~AU) and 2013 May 30 \citep[$R_h=6.87$~AU;][]{kulyk} yielded a non-detection of coma. Our observations on 2017 Aug 21 likewise did not show signs of coma material and provided the first color measurements of this object during a cessation of activity, revealing a relatively neutral surface color consistent with gray Centaurs.

\subsection{P/2011 S1}\label{subsec:P2011}
P/2011 S1 was discovered on 2011 Sep 18 as an active Centaur with a relatively low-eccentricity ($e=0.2$) orbit:$q=6.9$ and $Q=10.4$. The discovery images were obtained when the object was 7.53~AU from the Sun \citep{gibbs2011}. A series of serendipitous observations of P/2011 S1 were collected by the PAN-Starrs survey. Thirty observations, each comprised of several stacked individual exposures, were made from 2010 Sep 29 ($R_h=7.93$~AU) to 2012 Nov 4 ($R_h=7.15$~AU) along the inward branch of the object's orbit \citep{lin2014}. P/2011 S1 was active throughout these observations, though the level of the measured activity varied significantly --- from 40 to 150~kg/s, given similar assumptions for grain size and coma expansion velocities to the ones we make in this work. Subsequently, published observations from 2011 Sep 5 ($R_h=7.54$~AU) and 2013 Feb 7 ($R_h=7.09$~AU) likewise revealed activity, as well as a color measurement of $V-R=0.51\pm 0.12$ --- consistent with the gray Centaur subpopulation \citep{kulyk}. The most recent observation previous to ours of this object took place on 2014 Feb 26 ($R_h=6.91$~AU) yielded the color measurements $B-V=0.96 \pm 0.11$  and $V-R=0.59 \pm 0.03$ \citep{jewitt2015}.

From our observations on 2017 Feb 16 ($R_h=7.36$~AU), after the object passed perihelion, we computed a similar estimate for the activity level: $140\pm20$~kg/s.  When accounting for the relatively small size of its nucleus \citep[$\sim$7~km in diameter;][]{lin2014}, P/2011 S1 is one of the most active Centaurs known, comparable to 29P/Schwassmann-Wachmann 1, 174P, and P/LG (also known as P/2010 TO20). A notable characteristic of P/2011 S1 revealed by annular aperture photometry is that its coma appears to be somewhat redder than the near-nucleus region \citep{lin2014,kulyk}, which is opposite the trend observed for 167P and 174P (third outburst only).

\subsection{C/2012 Q1}\label{subsec:C2012}
C/2012 Q1 has a high-eccentricity orbit ($e=0.64$) and a perihelion distance of $q=9.48$; with an orbital semi-major axis of $a=26.15$~AU, this object is the second most distant active Centaur known. C/2012 Q1 was discovered in images obtained from 2012 Aug 28 to 2012 Sep 4 ($R_h=9.5$~AU), when a coma was reported \citep{miller2012}. Our observations on 2017 Aug 21 ($R_h=12.79$~AU) likewise showed an extended coma (see Figure~\ref{fig:SPB}), from which we computed a comparatively high mass loss rate of $250\pm40$~kg/s.

A previous set of observations by \citet{jewitt2015} on 2012 Oct 13--14 ($R_h=9.55$~AU) and 2013 Oct 1 ($R_h=9.85$~AU) detected activity and produced color measurements at various aperture sizes. Figure~\ref{fig:colors} shows the color measurements for the smallest $2.7''$ diameter aperture --- $B-V=0.85\pm0.03$ and $V-R=0.54\pm0.02$. While the $B-V$ color measurement derived from our observations is redder, given the much larger uncertainties on our measurement, our value is formally consistent with the previous value at $1.3\sigma$. For the $5.4''$ diameter aperture from \citet{jewitt2015}'s 2013 Oct 1 observation, which corresponds to a spatial extent most comparable to the area around the nucleus subtended by our optimal aperture, the listed color values are $B-V =0.94 \pm 0.03$  and $V-R = 0.51 \pm 0.02$; this $B-V$ value is more consistent with our measurement (at better than $1\sigma$). \citet{jewitt2015} carried out a multi-aperture analysis of this object and found no significant variation in color with angular radius. We conducted a similar analysis and likewise did not find any color trend, albeit with significantly larger uncertainties.

We can derive an upper limit on the nucleus size from aperture photometry of the near-nucleus region. The nucleus cross section is given by an expression analogous to Equation~\eqref{eq:cross}:
\begin{equation}C_n\equiv \pi \left(\frac{D}{2}\right)^{2}= 2.25\times10^{22} \frac{\pi R^2\Delta^2 10^{-0.4(m-m_{\Sun})}}{p_{R}\Phi(\alpha)},\end{equation}
where $D$ is the nucleus' diameter, and we set $m$ to be the R-band apparent magnitude of the object listed in Table~\ref{tab:obs}. Using the same assumptions as before for albedo $p_{R}$ and phase function $\Phi(\alpha)$, we obtain an upper limit of $D=24$~km.

\section{Discussion}\label{sec:discussion}
We now discuss the implications of our results for the current understanding of Centaur activity and colors.

\subsection{Centaur activity}\label{subsec:activity}

The cause of Centaur activity has been addressed in many previous works in the context of both observations and modeling. The range of heliocentric distances at which activity has been observed on these objects extends beyond the range at which water ice is unstable to sublimation, and even at the closest heliocentric distances where active Centaurs approach perihelion, the mass loss rate from water ice sublimation is too low to account for the levels of observed activity \citep[e.g.,][]{jewitt}. Meanwhile, other ices abundant in the bulk composition of outer solar system small bodies such as CO and CO$_2$ are far too volatile and would sublimate readily throughout the giant planet region if exposed at or near the surface, inconsistent with the lack of reported activity beyond $\sim$15~AU. 

The leading hypothesis for the trigger of Centaur activity is the crystallization of amorphous ice \citep[see discussion in][and references therein]{jewitt2015}. Simple energy balance calculations \citep{jewitt} and more sophisticated thermal modeling with heat transport \citep{guilbert} have shown that complete or partial crystallization occurs on timescales comparable to the orbital stability lifetime of Centaurs at heliocentric distances smaller than $\sim$16~AU. The latter study also demonstrates the effect of obliquity on the crystallization, with activity at the farthest distances requiring very high obliquity. The highly porous nature of amorphous ice can lead to trapped volatile gases such as CO and CO$_2$ upon condensation from the vapor phase during planetesimal formation. When Centaurs are scattered inward from the Kuiper Belt region and attain sufficiently high surface temperatures to allow for crystallization to proceed, the trapped gases are released, expelling dust through gas drag in the process.

Our observations, along with the body of previously-published results, are nominally consistent with this picture of Centaur activity (see Figure~\ref{fig:activity}). The two Centaurs for which we detected activity were at 7.36 and 12.79~AU from the Sun at the time of observation --- within the range predicted from theory and modeling of activity driven by amorphous ice crystallization. In particular, the very distant activity of C/2012 Q1 --- one of the most distant instances hitherto observed --- can be explained by crystallization, provided that the object has a substantial obliquity. Previous modeling of 174P's coma morphology following its first distant outburst ($R_h\sim13$~AU) also indicated high obliquity \citep{rousselot2015}. Another notable observation is the cessation of discernible activity on 167P during the outbound branch of its orbit. As illustrated in the figure, this object was last reported to be active on 2012 Oct 13 at a heliocentric distance of 15.33~AU \citep{jewitt2015}. During our observations on both 2017 Feb 15 and 2017 Aug 21, when the object was at heliocentric distances of more than 17~AU, we did not detect any discernible activity. Together, these observations suggest that the activity on 167P ceased as the surface cooled to below the temperature at which crystallization occurs. 

While the model of amorphous ice crystallization provides an explanation for Centaur activity in general, this process alone does not account for the trends in observed activity. The documented activity history of Centaurs demonstrates that the activity on at least some objects appears to be sporadic and highly variable, even when the objects are situated within the heliocentric range for which the crystallization process is expected to be active. Besides 174P, P/2008 CL94 is an illustrative example: although this object has a close and relatively circular orbit, it has been observed to become dormant, including during our observations. We also recall the significant activity level variations previously observed on P/2011 S1 \citep[see Section~\ref{subsec:P2011};][]{lin2014}.

\begin{figure*}[t!]
\centering
\includegraphics[width=\linewidth]{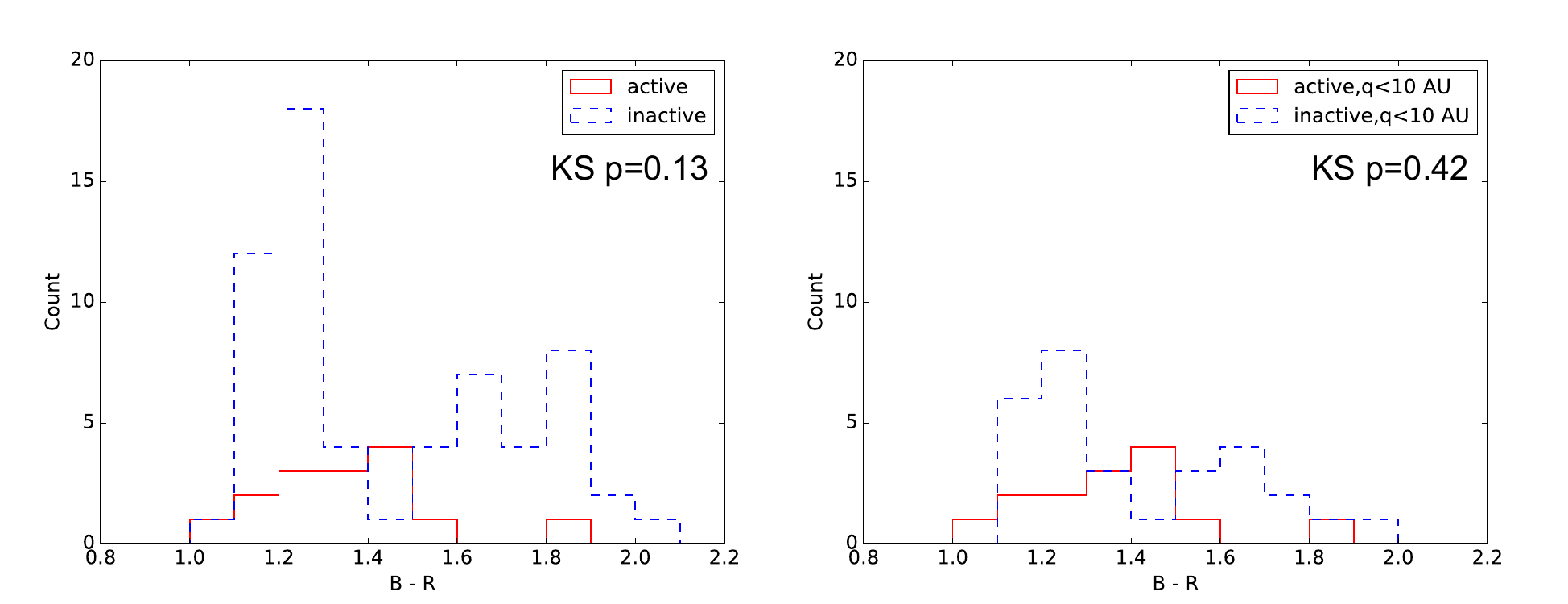}
\caption{Left: histogram of $B-R$ colors of active and inactive Centaurs  \citep[][and this work]{jewitt,mazzotta2014,jewitt2015,peixinho2015,tegler2016,mazzotta}, along with the calculated p-value from a two-sample Kolmogorov-Smirnov test on the two color distributions. Right: same, but limited to objects with orbital perihelia less than 10~AU.}
\label{fig:histograms}
\end{figure*}

We provide a few speculative explanations for the sporadic nature of Centaur activity within the framework of the current amorphous-to-crystalline water ice model. First, there may be significant inhomogeneities in the relative amount of trapped volatile gases in the amorphous ice matrix. Since the visible activity is driven by gas production and the associated expulsion of dust particles, if there are variations in the amount of trapped gasses, either across the surface of the object or with depth, then the mass loss rate would modulate with time as the crystallization front proceeds from the surface down to the lower layers of the regolith. This explanation is also feasible for possible patches of bulk CO ice in the subsurface that may become accessible to drive activity as the crystallization front reaches those depths \citep[e.g.,][]{wierzchos}. 

A second related possibility involves inhomogeneities in the physical properties of the non-volatile regolith component. The range of latitudes for which the surface temperature rises above the crystallization temperature varies throughout a Centaur's orbit. If there are local variations in the porosity or thermal inertia of the surface regolith, which affects both heat transfer and gas permeability to the surface, then the level of activity may change as the depth and surficial extent of the active regions change. Detailed modeling of the coma morphology of both 29P and 174P during its first outburst indicated the presence of distinct localized regions of activity \citep{miles2016,rousselot2015}, which suggests that only small regions of the surface had the particular physical and chemical conditions necessary for outgassing.

Lastly, the consequences of activity itself may produce a sporadic temporal trend. When a Centaur is active, some of the expelled dust settles back down and blankets the surface with fine particulate matter. With time, a crust of fallback debris forms on the surface, which may severely reduce the permeability from the subsurface \citep[e.g.,][]{coradini2008}. As the thermal wave continues to penetrate the surface and the crystallization front proceeds, the released volatile gas molecules are unable to escape to the surface, and as a result, pressure builds underneath the surficial crust until explosive outgassing occurs, breaking through the surface layers and producing a strong outburst in activity. Such a scenario has been invoked to explain, for example, the periodic nature of activity outbursts on 29P and the distinct outbursts on 174P \citep[e.g.,][]{miles2016,wierzchos,explosion}.

Stepping back to assess the broader picture, it is important to note that while the crystallization of amorphous ice is generally consistent with the body of previous observations, no incontrovertible evidence of amorphous ice has hitherto been obtained for Centaurs, or among the population of Kuiper Belt objects from which Centaurs are sourced. Likewise, observational studies of both short- and long-period comets have not yielded any positive detections. In light of the absence of concrete spectroscopic evidence for the existence of significant amounts of amorphous ice on the surfaces of Centaurs and KBOs, the question of the cause of Centaur activity remains open. 

\subsection{Centaur colors}\label{subsec:colors}

The issue of the discrepant color distributions of active and inactive Centaurs was first pointed out in \citet{jewitt}. As described in Section~\ref{sec:results} and illustrated in Figure~\ref{fig:colors}, while the color distribution of inactive Centaurs displays a robust bimodality, the color distribution of active Centaurs is not discernibly bimodal, with a general lack of the redder colors characteristic of the red Centaur subpopulation. The recent photometric observations of 523676 \citep{mazzotta} revealed the first red active Centaur. In light of this, we revisit the subject of Centaur colors.

Since we are concerned with the surface color of the nuclei when addressing the color distribution of active Centaurs, the first question is whether the measurements of active Centaurs are affected by the presence of coma material. Several earlier works raised the possibility of a blueward color bias from coma particulates affecting the measured photometry of active Centaurs through non-geometric scattering off optically small dust grains \citep[e.g.,][]{jewitt}. However, multi-aperture photometry of active Centaurs as well as JFCs \citep[e.g.,][and our analysis of C/2012 Q1 in Section~3.5]{bauer2008,rousselot2008,mazzotta2014,jewitt2015} has demonstrated that while systematic trends in color with increasing angular radius do exist for some objects, these variations tend to be small, typically a few $\times$0.01~mag, which is significantly smaller than the characteristic difference in colors between gray and red Centaurs. Moreover, these trends in spectral gradient can be both positive and negative, and in most cases, the observations are statistically consistent with no variation with angular radius, given the measurement uncertainties.

There are exceptions documented in the literature, e.g., 167P \citep{jewitt2015}, 174P \citep[third outburst only;][]{seccull}, and 523676 \citep{mazzotta}, for which comparative photometry of the near-nucleus region and pure coma revealed comae that were significantly bluer than the nucleus. Nevertheless, looking at the bigger picture, it does not appear that geometric effects impart any significant or systematic bias to the measured colors of Centaurs during periods of activity. This conclusion is strongly supported by measurements of several dormant active Centaurs \citep[e.g.,][]{bauer2003,kulyk,seccull}, including the ones described in this paper for 167P, 174P, and C/2008 CL94. Comparisons of the colors obtained during activity and dormancy show that the values are generally consistent. Furthermore, \citet{jewitt2015} demonstrated that while optically small particles are abundant in the coma material, the scattering cross section is still predicted to be dominated by larger particles, which suggests that blue scattering is not generally expected.

Having argued against systematic bias in the measured colors of active Centaurs due to blue scattering off small coma particles, we now proceed to a statistical analysis of the Centaur color distribution. The left panel in Figure~\ref{fig:histograms} shows the $B-R$ color distribution for active and inactive Centaurs. Carrying out a non-parametric two-sample Kolmogorov-Smirnov (KS) test on these two color distributions, we obtain a p-value of 0.13, indicating that the chance of the two distributions being drawn from the same underlying distribution is 13\%. For a Gaussian distribution, this value corresponds to a roughly $1.5\sigma$ difference, which is lower than the roughly $2\sigma$ difference reported by \citet{jewitt2015} prior to the discovery of the red active Centaur 523676.

In the previous subsection, we described the prevailing hypothesis that ascribes the cause of Centaur activity to the crystallization of amorphous ice. Since this process is only effective at relatively small heliocentric distances, a more diagnostic statistical analysis is to compare the color distributions of active and inactive Centaurs with orbits that would allow for activity to occur. In the right panel of Figure~\ref{fig:histograms}, we plot the color distributions of active and inactive Centaurs, limiting the samples to objects with perihelion distances less than 10~AU. This arbitrary cutoff includes all active Centaurs except 167P. Varying the cutoff distance by several AU does not significantly affect the results. Running the KS test on these low-perihelion color distributions returns $p=0.42$, which corresponds to a $0.8\sigma$ difference. Therefore, we conclude that the color distribution of active Centaurs does not differ from that of inactive Centaurs with similar orbits in a statistically significant way.

The newly demonstrated statistical consistency of the active and inactive Centaur color distributions strongly hinges upon the addition of the red Centaur 523676 to the sample of active objects. It has been proposed that blanketing of the strongly irradiated regolith by unirradiated fallback material from the ejected coma could be responsible for quickly converting initially red Centaurs to gray objects upon the onset of activity \citep[e.g.,][]{jewitt,jewitt2015}. \citet{mazzotta} report that the coma of 523676 is significantly more neutral than the near-nucleus region and is consistent with the color of gray Centaurs. They invoke the blanketing hypothesis in suggesting that the blanketing process on that Centaur must be incipient. Order of magnitude estimates for the characteristic blanketing timescale for such an object are very short, ranging from tens to hundreds years, even given the relatively low calculated low mass loss rate of $\sim$10~kg/s. Assuming a typical Centaur lifetime of $10^{6-7}$~years, this means that 523676 must have been observed within the first $10^{-6}$--$10^{-4}$ fraction of its lifetime --- comparable to a few orbital timescales.

The blanketing scenario predicts that all Centaurs, regardless of their primordial color, become gray once activity initiates. It follows that the active Centaur color distribution should be unimodal and therefore distinct from the inactive Centaur color distribution, with 523676 representing an exceptional object that was fortuitously observed almost immediately after the first onset of activity.

While we cannot rule out this scenario, we also cannot categorically rule out the possibility that geometric effects may be the explanation for the significantly more neutral coma of 523676. In other documented cases of significantly bluer comae, such as 167P and 174P (see Sections~\ref{subsec:167P} and \ref{subsec:174P}, there were epochs during which relatively blue coma colors were measured, as well as observations at other times that indicated no significant color difference between coma and nucleus. This suggests that the instances of relatively blue coma may not have been representative of the intrinsic color of the dust, but rather contingent upon some particular aspect(s) of the coma morphology during those times. Future observations of 523676 might likewise show a coma color that is more consistent with the red color of the nucleus.

\subsubsection{The possible role of thermal processing}\label{subsubsec:hypothesis}

Another dimension in the discussion of Centaur colors is orbital history. In Section~\ref{subsec:activity}, we described how the observed activity history of Centaurs is consistent with amorphous ice crystallization being the primary driver of activity. This temperature-driven process is only effective at heliocentric distances less than $\sim$16~AU \citep{guilbert}. Naturally, this means that the average perihelion distance of active Centaurs is smaller than the overall Centaur population, as well as the inactive Centaur population, a fact that has been demonstrated by several previous works \citep[e.g.,][]{jewitt,jewitt2015}. In particular, dynamical simulations of Centaurs have shown that, based on their current orbits, active Centaurs spend on average a significantly larger fraction of time at smaller heliocentric distances than inactive Centaurs \citep{melita,model2}. Notably, these same simulations also show that \textit{both inactive and active} gray Centaurs have statistically closer orbits than red Centaurs.

Since the close encounters that scatter KBOs inward onto Centaur orbits are random, there is no reason that gray KBOs would be preferentially scattered onto lower-perihelion orbits than red KBOs. It follows that some secondary process must be responsible for neutralizing the surface material on red Centaurs with low-perihelion orbits, thereby producing the apparently divergent orbital distributions of gray and red Centaurs. Within the blanketing hypothesis, this discrepancy would be due to the enrichment of gray objects in the population of inactive low-perihelion Centaurs as a consequence of activity: red objects that have low perihelia become active, and the activity rapidly blankets their surfaces with unweathered, relatively-neutral dust, before the object eventually becomes permanently dormant. Given the recent observation of a red active Centaur, as well as several low-perihelion inactive red Centaurs with orbits that would allow for activity (see right panel of Figure~\ref{fig:histograms}), we propose instead that thermal processing of the surface may be the cause of red-to-gray Centaur conversion.

The scenario is as follows: gray and red KBOs are scattered into the giant planet regions randomly, and initially, the two color subpopulations are distributed evenly in the space of orbital semi-major axis and perihelion distance. Some of the Centaurs (both gray and red) that have closer-in orbits become active, though the activity does not directly affect the surface colors. Over time, objects that have closer orbits accumulate higher levels of irradiation and heating and become gray, even if activity never occurs and the objects remain inactive. Eventually, the surface color of all red Centaurs reaching a certain threshold of thermal processing is neutralized, with only the more distant red Centaurs retaining their primordial color over long timescales. 

For another perspective on this hypothesis, we turn to JFCs. Like the Centaurs, JFCs are also former KBOs that have been scattered inward, but their orbits lie much closer to the Sun, with perihelion distances near or below that of Jupiter. The color distribution of JFCs differs dramatically from that of Centaurs, lacking red objects entirely \citep{jewitt2015}. In the context of our hypothesis, the reason for this is straightforward --- all of the JFCs have undergone vigorous heating and irradiation, regardless of whether cometary activity ever occurred, and the initially red objects have rapidly neutralized at the significantly higher temperatures of the inner Solar System.

The general lack of robust constraints on the detailed composition of Centaurs means that the qualitative hypothesis outlined above is necessarily speculative. In particular, the relevant timescale over which the thermal processing would occur is uncertain, though it could in principle be longer than the blanketing timescale for active Centaurs, thereby allowing for red active Centaurs to retain their color for longer than a few orbital periods. A discovery of another red active Centaur, for example, would strongly disfavor the blanketing explanation for the neutralization of Centaur colors, since the chances of observing two objects within the first 10--100~years of their activity lifetimes are astronomically small, as explained previously.

We mention in passing that models of KBOs, and by extension, Centaurs predict that the surfaces of these objects are composed of a refractory irradiation mantle from the weathering of primordial volatile ices in the outer Solar System \citep[e.g.,][]{brownschaller,wong3}. Previous experimental work has shown that irradiation of volatile ices leads to a general reddening of the visible spectral gradient to values typical for KBOs and the production of complex hydrocarbons \citep[e.g.,][]{brunetto}. Laboratory studies of similar high molecular weight hydrocarbons, such as asphaltite and kerite, as well as direct irradiation products of volatile ice mixtures, have demonstrated the neutralization of the optical color under irradiation, due to carbonization of the surface \citep[e.g.,][]{moroz,poston}. Complex organics similar to the ones analyzed in irradiation studies have been detected on the surface of the highly-active JFC 67P/Churyumov-Gerasimenko \citep[e.g.,][]{wright}, so we consider the aforementioned hypothesis of the neutralizing effect of thermal processing to be at least plausible.

\section{Conclusion}
We have presented the results of multiband photometric observations of Centaurs: five active Centaurs --- 167P/CINEOS, 174P/Echeclus, P/2008 CL94, P/2011 S1,  C/2012 Q1, and four inactive red Centaurs  --- 83982 Crantor, 121725 Aphidas, 250112 2002 KY14, 281371 2008 FC76. 

\begin{itemize}
\item Activity was detected on two Centaurs --- P/2011 S1 and C/2012 Q1. Estimation of the activity level using similar assumptions to previous works in the literature yielded relatively high mass loss rates of $140\pm20$ and $250\pm40$~kg/s, respectively.
\item No discernible comae were found on the active Centaurs 167P/CINEOS, 174P/Echeclus, and P/2008 CL94 at the time of our observations. The measured $B-V$ and $V-R$ colors of these Centaurs during inactivity are consistent with the colors of the gray Centaur subpopulation and statistically identical to previously-published colors obtained during periods of activity.
\item We did not detect the onset of activity on the four inactive red Centaurs in our sample.
\item The observed activity history of active Centaurs continues to be consistent with the hypothesis that crystallization of amorphous water ice is the primary trigger.
\item Comparisons of measured colors of active Centaurs obtained during periods of activity and dormancy indicate that photometric observations of active Centaurs are generally not affected by significant contamination from blue scattering off coma particulates.
\item With the discovery of the red active Centaur 523676 \citep{mazzotta}, the color distributions of active Centaurs and inactive Centaurs with similar orbits are now statistically consistent to within $1\sigma$.
\item We propose that the higher levels of thermal processing experienced by lower-perihelion Centaurs, both active and inactive, might lead to neutralization of their surface colors, resulting in the observed divergence in perihelion distance distributions between gray and red Centaurs.
\end{itemize}

\acknowledgements
I.W. is supported by a Heising-Simons Foundation \textit{51 Pegasi b} postdoctoral fellowship. A.M. thanks the Research Internship in Science and Engineering program organized by Boston University for the summer research opportunity that contributed substantially to this work. This work made use of the JPL Solar System Dynamics high-precision ephemerides through the HORIZONS system.

\end{document}